\documentclass[letterpaper,twocolumn]{jpsj3}
\usepackage{txfonts}
\usepackage{amsmath}
\newcommand{\bvec}[1]{\mbox{\boldmath $#1$}}

\title{Superconductivity enhanced by abundant low-energy phonons in (Sr$_{1-x}$Ca$_x$)$_3$Rh$_4$Sn$_{13}$}

\author{Yasuhiro Terasaki, Ratsuki Yamaguchi, Yui Ishii\thanks{ishii@mtr.osakafu-u.ac.jp}, Yurina Tada, Arisa Yamamoto, and Shigeo Mori}
\inst{Department of Materials Science, Osaka Prefecture University, Sakai, Osaka 599-8531, Japan} 

\abst{
The effects of structural quantum criticality on the strong-coupling superconductivity of (Sr$_{1-x}$Ca$_x$)$_3$Rh$_4$Sn$_{13}$ have been investigated via electrical resistivity and specific heat measurements.
We demonstrate that the lattice specific heat at low temperatures considerably increases toward the structural quantum critical point (SQCP), $x_{\rm c}\approx0.9$.
The superconducting gap increases with $x$ in the exact same fashion as the low-temperature lattice specific heat, clearly indicating that the abundant low-energy phonons cause strong-coupling superconductivity.
Despite the electron-phonon interaction, which is much more enhanced than the electron correlation, the low-temperature electrical resistivity near the SQCP varies as $\propto T^2$.
This finding suggests that structural quantum fluctuation affects the power law arising from the electron-phonon scattering.
}

\begin{document}
\maketitle

Competition among the degrees of freedom of electrons and a lattice has long been a central issue in solid-state physics.
Recent interest has also been directed to phonon instability and superconductivity enhanced near the structural quantum critical point (SQCP), at which condensation of a soft mode is suppressed down to the absolute zero temperature\cite{MoTe2,LaCuAu,SF3}.
A soft mode is a phonon whose frequency decreases upon cooling. 
When the soft-mode frequency eventually reaches zero, a structural phase transition occurs in a manner that reflects the atomic vibration pattern of the soft mode.
In most cases, the structural phase transition associated with a soft mode is of a second order.
A charge density wave (CDW) transition is one of the examples that involves phonon softening as a pretransitional phenomenon.
The second-order Jahn-Teller effect is also known as the microscopic origin of soft modes\cite{2JT}.

Historically, studies on structural quantum criticality began with the discovery of quantum paraelectrics in the 1970s \cite{STO_quantumpara,KTO_quantumpara}. 
These materials are insulating but have provided us with essential information for exploring the quantum effects in phonon dynamics.
SrTiO$_3$ is a well-known quantum paraelectric that is immediately next to the SQCP.
Although it shows no ferroelectric phase transition by itself down to the absolute zero temperature, it has a ferroelectric soft mode in one of the optical branches\cite{STO_Eu}.
The ferroelectric phase emerges by applying physical pressure\cite{STO_stress} or chemical substitution with Ca and $^{18}$O \cite{STO_Ca,18Oexchanged}.
Importantly, outside the ferroelectric phase, phonon softening is aborted at $\approx$10 K, below which the soft-mode frequency exhibits an almost constant value \cite{Yamada_JPSJ26,TaniguchiPRB72,TaniguchiPRL99}.
In addition, superconductivity and ferroelectricity coexist near the SQCP of dilute carrier-doped Sr$_{1-x}$Ca$_x$TiO$_{3-\delta}$\cite{STO_super4,Ferroelectric_Super}, which has recently accelerated studies on structural quantum criticality from both experimental and theoretical viewpoints\cite{Coak,Setty,Kanasugi,Edge}.
It is crucial to clarify how lattice instability near the SQCP affects the electron-phonon interaction as well as superconductivity.

The ternary stannide $R_3T_4$Sn$_{13}$ ($R$ = Ca, Sr, La; $T$ = Co, Rh, Ir) is one of the best systems to investigate the nature of the SQCP\cite{Akimitsu,La3Co4Sn13_1,La3Co4Sn13_2,CaIrSn_JPSJ79,Wang_PRB86,Zhou_PRB86,Gerber_PRB88,Biswas_PRB90}. 
Most of these materials exhibit both superconductivity and a structural phase transition.
They have a common high-temperature structure (a space group $Pm\bar{3}n$) that comprises Sn(2)$_{12}$ icosahedra packed in a CsCl arrangement. 
Each icosahedron includes Sn(1) as a guest atom and forms a three-dimensional framework by sharing its plane with a $T$Sn$_6$ trigonal prism.
The cavities formed by the framework are occupied by $R$ atoms.
The microscopic origin of the structural phase transition remains unclear\cite{Suyama_PRB97}, although CDW instability has been proposed for some of these compounds\cite{Goh_PRL109,Kuo_PRB89,Fang_PRB90,Kuo_PRB91}.

The focus of this study is a solid solution system, (Sr$_{1-x}$Ca$_x$)$_3$Rh$_4$Sn$_{13}$, which exhibits an SQCP at $x_{\rm c}\approx0.9$\cite{Goh_PRL115}.
The parent compound Sr$_3$Rh$_4$Sn$_{13}$ is regarded as a conventional $s$-wave superconductor with $T_{\rm c}$ = 4.2 K\cite{Akimitsu}.
This system experiences a structural phase transition at $T_{\rm s}$= 138 K, which results in a low-temperature structure ($I\bar{4}3d$) with an enlarged cell size of $2a \times 2b \times 2c$\cite{Goh_PRL114}.
Studies using inelastic X-ray scattering and first-principles calculations have revealed that (Sr$_{1-x}$Ca$_x$)$_3$Rh$_4$Sn$_{13}$ has a soft mode with $\bvec{q}$=(0.5, 0.5, 0) in one of the acoustic branches, whose mode is responsible for the structural phase transition\cite{Goh_PRL114,Goh_PRB95,Goh_PRB93,Goh_PRB98}.
In the structure and dynamics studies of Ca$_3$Ir$_4$Sn$_{13}$, of which the high-temperature phase is isostructural with the Rh system, the soft mode has been characterized as a ``breathing'' mode of the Sn(2)$_{12}$ cage\cite{Mazzone_PRB92}.

While $T_{\rm s}$ is suppressed with increasing $x$ and disappears at the SQCP, $T_{\rm c}$ gradually increases with $x$ and reaches 8 K at $x=1$. 
Several intriguing properties have been reported near the SQCP\cite{Goh_PRL115}. 
For example, $2\Delta / k_{\rm B}T_{\rm c}$ exhibits a large value beyond the BCS value, which indicates strong-coupling superconductivity.
A large enhancement in the Sommerfeld coefficient, $\gamma$, has also been reported. 
In addition, the Debye temperature, $\Theta_{\rm D}$, abruptly drops at the SQCP.
Ca$_3$Rh$_4$Sn$_{13}$, which is in the vicinity of the SQCP, exhibits a glasslike lattice thermal conductivity\cite{Kuo_SSC}, although those of Sr$_3$Rh$_4$Sn$_{13}$ and Sr$_3$Ir$_4$Sn$_{13}$ behave as expected for crystalline solids\cite{Kuo_PRB89,Kuo_PRB91}. 
Specifically, the temperature dependence of the thermal conductivity gradually changes from crystalline behavior to amorphous behavior with increasing $x$\cite{Thesis}.
Furthermore, inelastic X-ray scattering experiments of (Sr$_{1-x}$Ca$_x$)$_3$Rh$_4$Sn$_{13}$\cite{Goh_PRB98} have revealed that the soft mode exists at $x_{\rm c}$ and survives down to approximately absolute zero temperature.
It is naturally anticipated that structural quantum fluctuation predominates the lattice dynamics near the SQCP, which has motivated us to investigate the lattice specific heat together with the electrical resistivity of (Sr$_{1-x}$Ca$_x$)$_3$Rh$_4$Sn$_{13}$ in expectation of a strong interaction between low-energy phonons and itinerant electrons.
In this paper, we demonstrate clear evidence that the strong-coupling superconductivity observed in the vicinity of SQCP is attributable to abundant low-energy phonons.

\begin{figure}[t]
\begin{center}
\includegraphics[width=80mm]{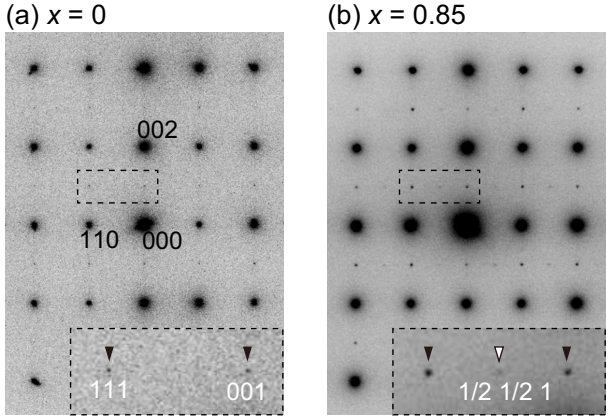}
\caption{\label{TEM} 
Electron diffraction patterns for (a) $x=0$ and (b) 0.85 obtained at room temperature with [$1\bar{1}0$] incidence. 
Enlarged views of the broken boxes are displayed in each inset.
Forbidden reflections with $00l$ and $hhl$ ($l\neq2n$) are indicated by closed triangles. 
As indicated by an open triangle, weak scattering is observed for $x=0.85$ at the 1/2 1/2 1 reciprocal point.
}
\end{center}
\end{figure}

Single crystals of (Sr$_{1-x}$Ca$_x$)$_3$Rh$_4$Sn$_{13}$ were grown using a tin-flux method\cite{Akimitsu}. 
As starting materials, bulk Sr (99\%), Ca granules (99.5\%), Rh wire (99.9\%), and Sn shots (99.99\%) were used. 
A mixture of them at a ratio of Sr : Ca : Rh : Sn = $2(1-x_{\rm n})$ : $2x_{\rm n}$ : 1 : 30 was placed in an alumina crucible and sealed in an evacuated silica tube.
After heating at 1050 $^{\circ}$C for 12 h, the tube was slowly cooled to 600 $^{\circ}$C at a rate of 2.5 $^{\circ}$C/h, followed by furnace cooling to room temperature.
The product was spun in a centrifuge and then immersed in hydrochloric acid to remove the Sn flux. 
The actual Ca concentration, $x$, in a chemical formula (Sr$_{1-x}$Ca$_x$)$_3$Rh$_4$Sn$_{13}$ was determined by using the inductively coupled plasma method.
The variation in lattice parameters evaluated utilizing powder X-ray diffraction on crushed single crystals follows Vegard's law, as reported previously\cite{Goh_PRL114}.
Electron diffraction was performed in a transmission electron microscope (JEOL, 2010M) at room temperature.
Electrical resistivity was measured by the four-probe method with alternating currents of 200--500 $\mu$A and a frequency of 314.16 Hz. 
For the resistivity measurements, small crystals were carefully thinned to obtain a good signal/noise ratio.
The heat capacity was measured by the heat-relaxation method in a physical property measurement system (Quantum Design).

Figures 1(a) and (b) represent electron diffraction patterns of the $x=0$ and 0.85 samples, respectively, obtained at 300 K with [$1\bar{1}0$] incidence.
The indices are based on the unit cell of the high-temperature phase.
Although the reported space group $Pm\bar{3}n$ has extinction rules for $00l$ and $hhl$ ($l\neq2n$) reflections, the $x=0$ sample exhibits a weak intensity at these reciprocal points, as indicated in the inset of Fig. 1(a) by filled triangles.
In addition, the intensity for $x=0.85$ is much stronger than that for $x=0$, as shown in Fig. 1(b).
The appearance of these reflections indicates symmetry reduction from the reported space group or local symmetry breaking in the $x=0.85$ sample.
Notably, a weak intensity is also observed for $x=0.85$ at the 1/2 1/2 1 reciprocal point, as indicated by an open triangle in the inset of Fig. 1(b), whose position corresponds to the M point of the Brillouin zone.
Because the soft mode has been confirmed to exist at the M point for $x=0.85$\cite{Goh_PRB98}, the observed diffuse-like weak intensity is ascribed to the soft mode.

\begin{figure}[t]
\begin{center}
\includegraphics[width=85mm]{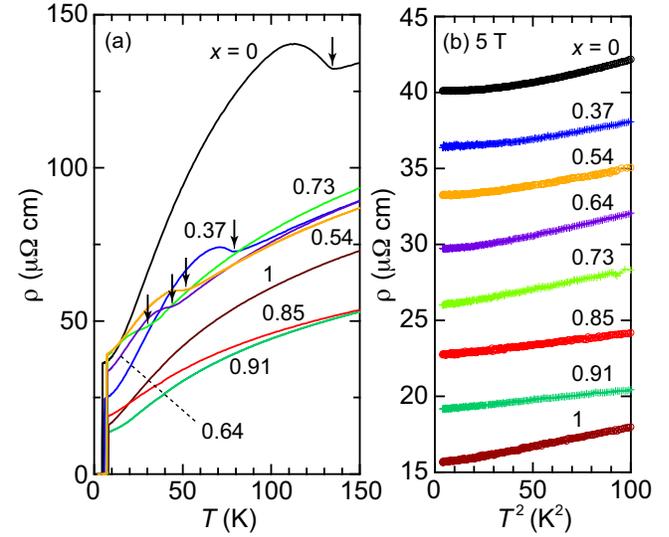}
\caption{\label{PhaseDiagram} 
(a) Electrical resistivity curves of (Sr$_{1-x}$Ca$_x$)$_3$Rh$_4$Sn$_{13}$ ($x=0\sim1$) in a zero external magnetic field. Enlarged view near the $T_{\rm c}$ is shown in Fig. S1(b)\cite{Supple}. The structural transition temperatures $T_{\rm s}$ are indicated by arrows. As $T_{\rm s}$ is decreased by increasing the Ca concentration $x$, the resistivity jump is also suppressed.
(b) Temperature variation in electrical resistivity in an external magnetic field of $\mu_0 H$ = 5 T. 
Data of $x=0.37$, 0.54, 0.64, 0.73, 0.85, and 0.91 are offset vertically by $+11$, $-6$, $-3.5$, $-12$, $+4.5$, and $+6$ $\mu \Omega$ cm, respectively.
}
\end{center}
\end{figure}

Sample qualities were verified via electrical resistivity and specific heat measurements in a zero external magnetic field.
Figure 2(a) shows the temperature dependence of the electrical resistivity, $\rho$, of $0\leq x \leq 1$ samples with a sharp superconducting transition at $T_{\rm c}$. 
In the resistivity curves, $T_{\rm c}$ is defined as the temperature at which zero resistivity was observed (Fig. S1)\cite{Supple}.
The $x = 0$ sample exhibits the structural phase transition at $T_{\rm s}$ = 135 K with a clear jump of $\rho$, as indicated by an arrow, which is consistent with previous reports.
$T_{\rm s}$ decreases with increasing $x$, and simultaneously, the resistivity jump becomes small.
The structural phase transition disappears at $x=0.85$, as reported. 
The residual resistivity ratio, $\rho_{300 {\rm \; K}}/\rho_0$, of our samples is evaluated to be 3$\sim$6.
As shown later in Fig. 4(a), the observed $T_{\rm s}$ and $T_{\rm c}$ describe the reported phase diagram well\cite{Goh_PRL115}, indicating the good quality of the obtained crystals.
The superconducting transition is also confirmed in specific heat measurements, as shown in Fig. S2\cite{Supple}, where a clear jump in the specific heat is observed at $T_{\rm c}$ for all samples.
In the specific heat measurements, $T_{\rm c}$ is defined as the midpoint of the specific heat jump.

Electrical resistivity in the normal state was investigated in an external magnetic field of $\mu_0 H$ = 5 T, as shown in Fig. 2(b).
The data are plotted as a function of $T^2$ with vertical offsets for clarity.
Although concave behavior is observed for the $x\leq0.64$ samples, the resistivity varies as almost $\propto T^2$ for the $x\geq0.73$ samples.
Power fits using $\rho = \rho_0 + AT^{\alpha}$ reveal that $\alpha$ yields 3.7 at $x=0$, where $\rho_0$ and $A$ are the residual resistivity and coefficient, respectively.
The value of $\alpha$ gradually decreases as $x$ increases and reaches 2.1 at $x=0.91$ with $A=1.1\times10^{-2} \; \mu \Omega$ cm K$^{-2}$.

To analyze the lattice contribution to the specific heat, the temperature variation in the total specific heat, $C$, in the normal state is plotted in the form $C/T^3$, as shown in Fig. 3(a).
Data below 10 K are obtained in an external magnetic field of $\mu_0 H$ = 6 T.
As indicated by an arrow in the curve of $x=0$, a broad peak is observed at $\approx$10 K, followed by a rapid increase upon cooling owing to the electronic specific heat, $\gamma T$, where $\gamma$ is the Sommerfeld coefficient.
The broad peak gradually shifts toward low temperatures as $x$ increases. 
Fig. 3 (b) represents the lattice specific heat, $C_{\rm lat}$, obtained by subtracting $\gamma T$ from the total specific heat. 
$\gamma$ is evaluated from the intercept of the $C/T-T^2$ plot shown in the inset. 
In general, the broad peak near 10 K arises from the excitation of all modes that depart from the Debye model.
The maximum value of the broad peak, ($C_{\rm lat}/T^3$)$_{\rm max}$, increases as $x$ increases and is almost invariant above $x = 0.85$. 
Concurrently, the value of $C_{\rm lat}/T^3$ at the lowest temperature significantly increases as well.
This result indicates that the phonon density of states, $F(\omega)$, considerably increases in the low-energy region below $\approx$ 1 meV.

\begin{figure}[t]
\begin{center}
\includegraphics[width=85mm]{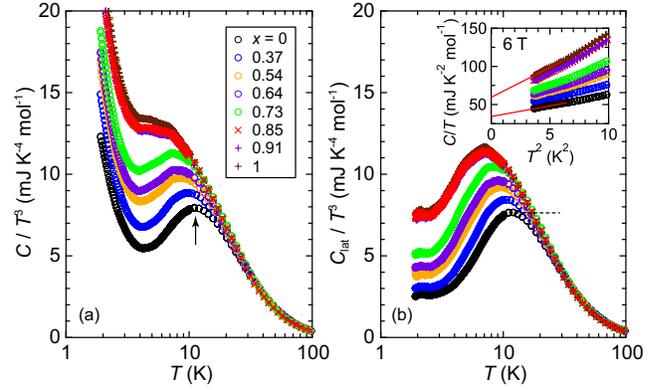}
\caption{\label{PhaseDiagram} 
(a) $C/T^3$ in the normal state plotted as a function of temperature. The horizontal axis is on a logarithmic scale. Data below 10 K are obtained in an external magnetic field of $\mu_0 H$ = 6 T.
For the $x=0$ sample, a broad peak is observed at $\approx$10 K as indicated by an arrow.
(b) Lattice specific heat, $C_{\rm lat}$, obtained by subtracting the electronic specific heat in the normal state, $\gamma T$, from the total heat capacity.
Inset shows the $C/T$ vs. $T^2$ plot. 
The sommerfeld coefficient, $\gamma$, and Debye temperature, $\Theta_{\rm D}$, are obtained by the linear fit, $C/T = \beta T^2 + \gamma$.
The maximum value of $C_{\rm lat}/T^3$ is defined by a broken line indicated in the figure.
}
\end{center}
\end{figure}

Figure 4 summarizes the parameters obtained from the resistivity and heat capacity data.
Fig. 4(a) represents $T_{\rm c}$ and $T_{\rm s}$ (left axis), together with the normalized superconducting gap, $2\Delta / k_{\rm B} T_{\rm c}$ (right axis).
The superconducting gap, $\Delta$, is obtained from the electronic specific heat in the superconducting state, $C_{\rm es}$, via a slope of the $\ln(C_{\rm es})$ vs. $T^{-1}$ plot (Fig. S3)\cite{Supple}.
As reported previously, $C_{\rm es}$ is fit well by the Arrhenius type $\exp (-\Delta/k_{\rm B} T)$ for the $0\leq x\leq 1$ samples, indicating the existence of an $s$-wave superconducting gap for all compositions \cite{Goh_PRL115}.
In Fig. 4(a), the value of $2\Delta / k_{\rm B} T_{\rm c}$ of $x=0$ exhibits a BCS value, $\approx 3.52$, indicating that Sr$_3$Rh$_4$Sn$_{13}$ is described as a fully gapped $s$-wave superconductor.
The value is rapidly elevated beyond the BCS value and reaches $\approx 5.5$ at $x=0.85$, indicating strong-coupling superconductivity in the vicinity of the SQCP \cite{Goh_PRL115}.
Fig. 4(b) represents the ($C_{\rm lat}/T^3$)$_{\rm max}$ (left red axis) and the $T^3$ coefficient, $\beta$, (right blue axis) at the lowest temperatures, both of which increase toward the SQCP.
Because the specific heat that varies as $\propto T^3$ at low temperatures arises from the long-wavelength acoustic modes with linear dispersion near the $\Gamma$ point, the increase in the coefficient $\beta$ indicates that the slope $\upsilon$ of the linear dispersion, $\omega = \upsilon q$, is reduced.
Near the SQCP compositions, there is the acoustic soft mode at the M point whose frequency approaches almost zero when $T\rightarrow0$, and this softening might have some effect on the slope of the linear dispersion near the $\Gamma$ point.
However, this effect is usually small. 
The increase in $\beta$ is thus mainly attributable to the decrease in $\Theta_{\rm D}$.
The variation in $\Theta_{\rm D}$ is depicted in Fig. 4 (b) together with the value of $\beta$.

\begin{figure}[t]
\begin{center}
\includegraphics[width=86mm]{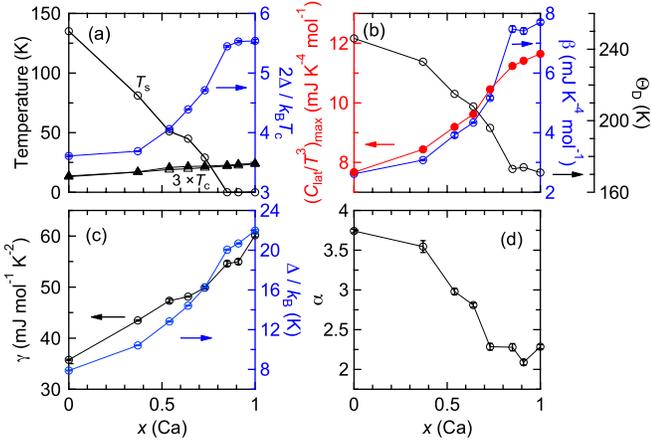}
\caption{\label{PhaseDiagram} 
(a) Structural phase transition temperature, $T_{\rm s}$, together with superconducting transition temperature, $T_{\rm c}$, (left axis) as a function of Ca concentration, $x$. 
As reported\cite{Goh_PRL115}, $T_{\rm s}$ decreases as $x$ increases and disappears at $x=0.85$.
The normalized superconducting gap, $2\Delta / k_{\rm B} T_{\rm c}$, is plotted on the right axis.
(b) The maximum value of $C_{\rm lat} / T^3$ (left axis) defined as the peak height in Fig. 2(d).
The $T^3$ coefficient $\beta$ obtained by the linear fit of the specific heat data is plotted on the right axis (blue) together with the Debye temperature, $\Theta_{\rm D}$ (black). 
($C_{\rm lat} / T^3$)$_{\rm max}$ and $\beta$ show a sharp increase toward the SQCP composition.
(c) Sommerfeld coefficient, $\gamma$, (left axis) together with the superconducting gap, $\Delta / k_{\rm B}$, determined by the specific heat measurements.
$\Delta / k_{\rm B}$ varies in the same fashion as ($C_{\rm lat} / T^3$)$_{\rm max}$.
(d) The exponent $\alpha$ in $\rho = \rho_0 + AT^{\alpha}$ obtained for the electrical resistivity in an external magnetic field of 5 T.
}
\end{center}
\end{figure}

On the other hand, $\gamma$ increases almost linearly with increasing $x$, as shown in Fig. 4(c) (left axis).
Notably, as shown in Fig. 4(c) (right axis), the superconducting gap $\Delta$ varies exactly in the same fashion as the $(C_{\rm lat}/T^3)_{\rm max}$.
Taking into account $F(\omega)$ in the electron-phonon interaction, the coupling constant, $\lambda$, of the interaction has the form $\lambda = 2\int^{\infty}_0 d\omega \;\alpha^2(\omega)F(\omega) / \omega$, where $\alpha^2(\omega)F(\omega)$ is the Eliashberg function. 
As shown in Fig. 4(b), $(C_{\rm lat}/T^3)_{\rm max}$ increases by a factor of 1.5 for $x=0\rightarrow1$. 
This leads to a rough estimation that the integral in the Eliashberg function would also increase, and correspondingly, the value of $\lambda$ as well.
The enhancement in $\Delta$ is rationalized by this increase in $\lambda$.
That is, the strong-coupling superconductivity observed near the SQCP stems from the additional vibrational states in the low-energy region.
Indeed, it has been reported that one of the acoustic modes with low energy can significantly contribute to the electron-phonon interaction.\cite{WTe2}

Figure 4 (d) represents the temperature exponent, $\alpha$, of the resistivity curves at the lowest temperatures, where $\alpha$ decreases with increasing $x$ and reaches $\alpha\approx 2$ near the SQCP. 
This fact might be apparently due to the strong electron-electron correlation that appears as the enhanced $\gamma$ shown in Fig. 4(c).
The value of $\alpha \approx 2$ was also reported for Ca$_3$Ir$_4$Sn$_{13}$\cite{CaIrSn_JPSJ79}.
However, this result conflicts with the decrease in $\Theta_{\rm D}$ observed in the specific heat measurements, as discussed below.
According to the well-known Bloch-Gr\"{u}neisen formula, the electrical resistivity arising from the electron-phonon interaction varies as 
\begin{equation}
\rho(T) \propto \left( \frac{T}{\Theta_{\rm D}} \right)^5 \int^{\frac{\Theta_{\rm D}}{T}} _0 
\frac{x^5}{(e^x-1)(1-e^{-x})} dx \; , \nonumber
\end{equation}
where the integral term yields an almost constant value for $T<\Theta_{\rm D}/15$.
Because the value of $\Theta_{\rm D}$ decreases to $\approx$170 K for $x\geq0.85$, as shown in Fig. 4(b), 
the integral term is regarded to be constant when $T<11$ K, which covers the temperature range where the $T^2$ dependence of $\rho$ is observed.
As shown in Fig. 4(b), $\Theta_{\rm D}$ decreases by a factor of 0.7 for $x=0\rightarrow1$, which gives rise to an increase in the temperature coefficient $( 1 / \Theta_{\rm D} )^5$ in the above formula by a factor of $\approx6$.
On the other hand, electron-electron interactions contribute to $\rho$ as $AT^2$, where the coefficient $A$ satisfies $A\propto\gamma^2$.
Because the value of $\gamma$ increases by a factor of 1.7 for $x=0\rightarrow1$ as observed in Fig. 4(c), the contribution from the electron correlation increases only by a factor of $\approx$3.
That is, the enhancement in the electron correlation is too small to overcome that in the electron-phonon interaction predicted by the Bloch-Gr\"{u}neisen formula.
This fact indicates that the contribution from electron-phonon scattering can be largely modified by the abundant phonons near the SQCP.

Because the increase in the ($C_{\rm lat}/T^3$)$_{\rm max}$ observed near the SQCP is so large, the single soft mode with low $\omega$ at the critical composition would not be the main factor for the abundant phonons.
Strong structural disorder likely exists in the vicinity of the SQCP as a result of the suppression of soft-mode condensation.
In structural phase transitions driven by a soft mode, structural instability is usually relieved by condensation of the soft mode.
The resultant low-temperature structure reflects the vibration pattern of the soft mode, as mentioned earlier.
Here, the vibration pattern of the soft mode of Sr$_3$Rh$_{14}$Sn$_{13}$ would essentially be the same as the ``breathing" mode of the Sn(2)$_{12}$ cages that has been reported for Sr$_3$Ir$_{14}$Sn$_{13}$\cite{Mazzone_PRB92} 
because the high-temperature phase of Sr$_3$Rh$_{14}$Sn$_{13}$ exhibits large anisotropic atomic displacement parameters in Sn(2) atoms\cite{Goh_PRL114}, which are very similar to those of Sr$_3$Ir$_{14}$Sn$_{13}$.
Sn(1) atoms exhibit only a small atomic displacement parameter, indicating that an additional contribution from a rattling is small.
The structural instability manifested itself as the soft mode is not relieved at the SQCP, and there is no development of a long-range superstructure\cite{Goh_PRB98}.
Nevertheless, the local structure realized near the SQCP should reflect the vibration pattern of the soft mode to some extent, giving rise to the Sn(2)$_{12}$ icosahedra deviating from the average structure, which should act as strong structural disorder.

Structural disorder can cause glasslike low-energy vibrational states\cite{Chumakov_PRL,halomethane,Baggioli}.
Considering the results of this study together with the previous reports, the thermal and structural properties of (Sr$_{1-x}$Ca$_x$)$_3$Rh$_{14}$Sn$_{13}$ near the SQCP can be summarized as follows.
First, this study reveals that the specific heat significantly increases in the low-temperature region particularly at the critical composition.  
Second, the temperature dependence of thermal conductivity transforms into that of amorphous solids as $x$ approaches the SQCP\cite{Kuo_SSC,Thesis}. 
Finally, the development of superlattice reflections is largely suppressed and the intensity becomes weak as the SQCP is approached\cite{Goh_PRB98}. 
These characteristics are very similar to those observed in Ba$_{1-x}$Sr$_x$Al$_2$O$_4$, which we have recently explored.
In Ba$_{1-x}$Sr$_x$Al$_2$O$_4$, the structural phase transition is caused by a ferroelectric soft mode and disappears near $x=0.07$\cite{Ishii_PRB93,Ishii_SciRep}.
As $x$ increases, the atomic displacement parameter of a particular atom increases systematically\cite{Kawaguchi_PRB}, and ferroelectric domains arising from the polar low-temperature structure are divided into small regions with a typical size of 10 nm\cite{Ishii_PRB94}.
The phonon spectrum at $x=0.07$ is reminiscent of a boson peak, where the low-energy phonons considerably increase\cite{Ishii_arXiv,Ishii_PRM}.
Similar phenomena may occur in the Sr$_3$Rh$_4$Sn$_{13}$ system.
The disordered atomic arrangement, which reflects the soft-mode vibration pattern within a short correlation length, is most likely to modify the lattice dynamics of the present system and affect the electron-phonon interactions near the SQCP.
It is a challenge in the future to clarify how the structural disorder affects the manner of electron-phonon scattering.
Studies on lattice dynamics including the phonon dispersion relation and local structure analyses are indispensable.

In summary, we have analyzed the electrical resistivity and specific heat of (Sr$_{1-x}$Ca$_x$)$_3$Rh$_4$Sn$_{13}$.
As the SQCP composition is approached, $(C_{\rm lat}/T^3)_{\rm max}$ at low temperatures significantly increases, indicating a large increase in $F(\hbar\omega)$ in the low-energy region.
This leads to an enhancement of electron-phonon interaction, resulting in strong-coupling superconductivity observed near the SQCP.
For electrical resistivity in the normal state, $T^2$ dependence is observed at low temperatures and near the SQCP composition, although the contribution from electron-phonon interactions should be largely enhanced. 
Atomic disorders caused by structural quantum fluctuations should affect the power law that arises from the electron-phonon scattering.

\begin{acknowledgment}

We thank Prof. A. Nakahira (Osaka Pref. Univ.) for his support in the compositional analysis using the inductively coupled plasma method.
This work was supported by a JSPS Grant-in-Aid for Scientific Research on Innovative Areas ``Mixed-anion'' (No. 17H05487, 19H04704) and JSPS KAKENHI (No. 20H01844).

\end{acknowledgment}


\begin{thebibliography}{99}
	\bibitem{MoTe2} H. Takahashi, T. Akiba, K. Imura, T. Shiino, K. Deguchi, N. K. Sato, H. Sakai, M. S. Bahramy, and S. Ishiwata, {\it Phys. Rev. B} {\bf 95}, 100501(R) (2017).
	\bibitem{LaCuAu} L. Poudel, A. F. May, M. R. Koehler, M. A. McGuire, S. Mukhopadhyay, S. Calder, R. E. Baumbach, R. Mukherjee, D. Sapkota, C. de la Cruz, D. J. Singh, D. Mandrus, and A. D. Christianson,  {\it Phys. Rev. Lett.} {\bf 117}, 235701 (2016).
	\bibitem{SF3} S. U. Handunkanda, E. B. Curry, V. Voronov, A. H. Said, G. G. Guzm\'{a}n-Verri, R. T. Brierley, P. B. Littlewood, J. N. Hancock, {\it Phys. Rev. B} {\bf 92}, 134101 (2015).
	\bibitem{2JT} I.B. Bersuker, {\it Phys. Lett.} {\bf 20}, 589 (1966).
	\bibitem{STO_quantumpara} K. A. M\"uller and H. Burkard, {\it Phys. Rev. B} {\bf 19}, 3593 (1979).
	\bibitem{KTO_quantumpara} G. A. Samara and B. Morosin, {\it Phys. Rev. B} {\bf 8}, 1256 (1973).
	\bibitem{STO_Eu} M. Takesada, M. Itoh, and T. Yagi, {\it Phys. Rev. Lett.} {\bf 96}, 227602 (2006).
	\bibitem{STO_stress} H. Uwe and T. Sakudo, {\it Phys. Rev. B} {\bf 13}, 271 (1976).
	\bibitem{STO_Ca} J. G. Bednorz and K. A. M\"{u}ller, {\it Phys. Rev. Lett.} {\bf 52}, 2289 (1984).
	\bibitem{18Oexchanged} R. Wang and M. Itoh, {\it Phys. Rev. B} {\bf 64}, 174104 (2001).
	\bibitem{Yamada_JPSJ26} Y. Yamada and G. Shirane, {\it Phys. Rev. Lett.} {\bf 26}, 396 (1969).
	\bibitem{TaniguchiPRB72} H. Taniguchi, T. Yagi, M. Takesada, and M. Itoh, {\it Phys. Rev. B} {\bf 72}, 064111 (2005).
	\bibitem{TaniguchiPRL99} H. Taniguchi, M. Itoh, and T. Yagi, {\it Phys. Rev. Lett.} {\bf 99}, 017602 (2007).
	\bibitem{STO_super4} B. S. de Lima, M. S. da Luz, F. S. Oliveira, L. M. S. Alves, C. A. M. dos Santos, F. Jomard, Y. Sidis, P. Bourges, S. Harms, C. P. Grams, J. Hemberger, X. Lin, B. Fauqu\'{e}, and K. Behnia, {\it Phys. Rev. B} {\bf 91}, 045108 (2015).
	\bibitem{Ferroelectric_Super} C. W. Rischau, X. Lin, C. P. Grams, D. Finck, S. Harms, J. Engelmayer, T. Lorenz, Y. Gallais, B. Fauqu\'{e}, J. Hemberger {\it et al.}, {\it Nat. Phys.} {\bf 13}, 643 (2017).
	\bibitem{Coak} M.J. Coak, C.R.S. Haines, C. Liu, S.E. Rowley, G.G. Lonzarich, and S.S. Saxena, {\it PNAS} {\bf 117}, 12707 (2020).
	\bibitem{Setty} C. Setty, {\it Phys. Rev. B} {\bf 99}, 144523 (2019).
	\bibitem{Kanasugi} S. Kanasugi and Y. Yanase, {\it Phys. Rev. B} {\bf 98}, 024521 (2018). 
	\bibitem{Edge} J. M. Edge, Y. Kedem, U. Aschauer, N. A. Spaldin, and A. V. Balatsky, {\it Phys. Rev. Lett.} {\bf 117}, 219901 (2016).
	\bibitem{Akimitsu} N. Kase, H. Hayamizu, and J. Akimitsu, {\it Phys. Rev. B} {\bf 83}, 184509 (2011). 
	\bibitem{La3Co4Sn13_1} Y. W. Cheung, J. Z. Zhang, J. Y. Zhu, W. C. Yu, Y. J. Hu, D. G. Wang, Y. Otomo, K. Iwasa, K. Kaneko, M. Imai, H. Kanagawa, K. Yoshimura, and S. K. Goh, {\it Phys. Rev. B} {\bf 93}, 241112(R) (2016). 
	\bibitem{La3Co4Sn13_2} A. \'{S}lebarski, M. Fija\l kowski, M. M. Ma\'{s}ka, M. Mierzejewski, B. D. White, and M. B. Maple, {\it Phys. Rev. B} {\bf 89}, 125111 (2014). 
	\bibitem{CaIrSn_JPSJ79} J. Yang, B. Chen, C. Michioka, and K. Yoshimura, {\it J. Phys. Soc. Jpn.} {\bf 79}, 113705 (2010).
	\bibitem{Wang_PRB86} K. Wang and C. Petrovic, {\it Phys. Rev. B} {\bf 86}, 024522 (2012). 
	\bibitem{Zhou_PRB86} S. Y. Zhou, H. Zhang, X. C. Hong, B. Y. Pan, X. Qiu, W. N. Dong, X.L. Li, and S.Y. Li, {\it Phys. Rev. B} {\bf 86}, 064504 (2012). 
	\bibitem{Gerber_PRB88} S. Gerber, J. L. Gavilano, M. Medarde, V. Pomjakushin, C. Baines, E. Pomjakushina, K. Conder, and M. Kenzelmann, {\it Phys. Rev. B} {\bf 88}, 104505 (2013). 
	\bibitem{Biswas_PRB90} P. K. Biswas, A. Amato, R. Khasanov, H. Luetkens, K. Wang, C. Petrovic, R. M. Cook, M. R. Lees, and E. Morenzoni, {\it Phys. Rev. B} {\bf 90}, 144505 (2014).
	\bibitem{Suyama_PRB97} K. Suyama, K. Iwasa, Y. Otomo, K. Tomiyasu, H. Sagayama, R. Sagayama, H. Nakao, R. Kumai, Y. Kitajima, F. Damay, J.-M. Mignot, A. Yamada, T. D. Matsuda, and Y. Aoki, {\it Phys. Rev. B} {\bf 97}, 235138 (2018).
	\bibitem{Goh_PRL109} L. E. Klintberg, S.K. Goh, P. L. Alireza, P.J. Saines, D. A. Tompsett, P. W. Logg, J. Yang, B. Chen, K. Yoshimura, and F. M. Grosche, {\it Phys. Rev. Lett.} {\bf 109}, 237008 (2012).
	\bibitem{Kuo_PRB89} C. N. Kuo, H. F. Liu, C. S. Lue, L. M. Wang, C. C. Chen, and Y. K. Kuo, {\it Phys. Rev. B} {\bf 89}, 094520 (2014).
	\bibitem{Fang_PRB90} A. F. Fang, X. B. Wang, P. Zheng, and N. L. Wang, {\it Phys. Rev. B} {\bf 90}, 035115 (2014).
	\bibitem{Kuo_PRB91} C. N. Kuo, C. W. Tseng, C. M. Wang, C. Y. Wang, Y. R. Chen, L. M. Wang, C. F. Lin, K. K. Wu, Y. K. Kuo, and C. S. Lue, {\it Phys. Rev. B} {\bf 91}, 165141 (2015).
	\bibitem{Goh_PRL115} W.C. Yu, Y.W. Cheung, P.J. Saines, M. Imai, T. Matsumoto, C. Michioka, K. Yoshimura, and S.K. Goh, {\it Phys. Rev. Lett.} {\bf 115}, 207003 (2015).
	\bibitem{Goh_PRB98} Y. W. Cheung, Y. J. Hu, M. Imai, Y. Tanioku, H. Kanagawa, J. Murakawa, K. Moriyama, W. Zhang, K. T. Lai, K. Yoshimura, F. M. Grosche, K. Kaneko, S. Tsutsui, and S. K. Goh, {\it Phys. Rev. B} {\bf 98}, 161103(R) (2018).
	\bibitem{Goh_PRB95} Y.J. Hu, Y.W. Cheung, W.C. Yu, M. Imai, H. Kanagawa, J. Murakawa, K. Yoshimura, and S.K. Goh, {\it Phys. Rev. B} {\bf 95}, 155142 (2017).
	\bibitem{Goh_PRL114} S.K. Goh, D.A. Tompsett, P.J. Saines, H.C. Chang, T. Matsumoto, M. Imai, K. Yoshimura, and F.M. Grosche, {\it Phys. Rev. Lett.} {\bf 114}, 097002 (2015).
	\bibitem{Goh_PRB93} Y. W. Cheung, J. Z. Zhang, J. Y. Zhu, W. C. Yu, Y. J. Hu, D. G. Wang, Yuka Otomo, Kazuaki Iwasa, Koji Kaneko, Masaki Imai, Hibiki Kanagawa, Kazuyoshi Yoshimura, and Swee K. Goh, {\it Phys. Rev. B} {\bf 93}, 241112 (2016).
	\bibitem{Mazzone_PRB92} D.G. Mazzone, S. Gerber, J.L. Gavilano, R. Sibille, M. Medarde, B. Delley, M. Ramakrishnan, M. Neugebauer, L.P. Regnault, D. Chernyshov, A. Piovano, T.M. Fern\'{a}ndez-D\'{i}az, L. Keller, A. Cervellino, E. Pomjakushina, K. Conder, and M. Kenzelmann, {\it Phys. Rev. B} {\bf 92}, 024101 (2015).
	\bibitem{Kuo_SSC} C. W. Tseng, C. N. Kuo, B. S. Li, L. M. Wang, A. A. Gippius, Y. K. Kuo, C. S. Lue, {\it Solid State Commun.} {\bf 270}, 26 (2018).
	\bibitem{Thesis} X. Chen, Ph. D thesis, University of Cambridge (2017).
	\bibitem{Supple} See Supplemental Material, which includes (i) AC resistivity curves near room temperature and $T_{\rm c}$, (ii) parameters obtained from the electrical resistivity and specific heat data, (iii) specific heat near $T_{\rm c}$, and (iv) electronic specific heat in the superconducting state.
	\bibitem{WTe2} P. Lu, J.-S. Kim, J. Yang, H. Gao, J. Wu, D. Shao, B. Li, D. Zhou, J. Sun, D. Akinwande, D. Xing, and J.-F. Lin, {\it Phys. Rev. B} {\bf 94}, 224512 (2016).
	\bibitem{Chumakov_PRL} A. I. Chumakov, G. Monaco, A. Monaco, W. A. Crichton, A. Bosak, R. R\"uffer, A. Meyer, F. Kargl, L. Comez, D. Fioretto {\it et al.}, {\it Phys. Rev. Lett.} {\bf 106}, 225501 (2011).
	\bibitem{halomethane} M. Moratalla, J. F. Gebbia, M. A. Ramos, L. C. Pardo, S. Mukhopadhyay, S. Rudi\'{c}, F. F.-Alonso, F. J. Bermejo, and J. L. Tamarit, {\it Phys. Rev. B} {\bf 99}, 024301 (2019).
	\bibitem{Baggioli} M. Baggioli and A. Zaccone, {\it J. Phys. Mater.} {\bf 3}, 015004 (2020).
	\bibitem{Ishii_PRB93} Y. Ishii, S. Mori, Y. Nakahira, C. Moriyoshi, J. Park, B. G. Kim, H. Moriwake, H. Taniguchi, and Y. Kuroiwa, {\it Phys. Rev. B} {\bf 93}, 134108 (2016).
	\bibitem{Ishii_SciRep} Y. Ishii, H. Tsukasaki, E. Tanaka, and S. Mori, {\it Sci. Rep.} {\bf 6}, 19154 (2016).
	\bibitem{Kawaguchi_PRB} S. Kawaguchi, Y. Ishii, E. Tanaka, H. Tsukasaki, Y. Kubota, and S. Mori, {\it Phys. Rev. B} {\bf 94}, 054117 (2016).
	\bibitem{Ishii_PRB94} Y. Ishii, H. Tsukasaki, E. Tanaka, S. Kawaguchi, and S. Mori, {\it Phys. Rev. B} {\bf 94}, 184106 (2016).
	\bibitem{Ishii_arXiv} Y. Ishii, A. Yamamoto, N. Sato, Y. Nambu, S. Ohira-Kawamura, N. Murai, T. Mori, and S. Mori, arXiv:2104.01969 [cond-mat.mtrl-sci].
	\bibitem{Ishii_PRM} Y. Ishii, Y. Ouchi, S. Kawaguchi, H. Ishibashi, Y. Kubota, and S. Mori, {\it Phys. Rev. Materials} {\bf 3}, 084414 (2019).

\end{thebibliography}
\end{document}


\begin{center}
Supplemental Material\\

\vspace{10mm}
{\large \bf Superconductivity enhanced by abundant low-energy phonons in (Sr$_{1-x}$Ca$_x$)$_3$Rh$_4$Sn$_{13}$}\\

\vspace{10mm}
Y. Terasaki, R. Yamaguchi, Y. Ishii$^*$, Y. Tada, A. Yamamoto, and S. Mori\\

\vspace{10mm}
Department of Materials Science, Osaka Prefecture University, Sakai, Osaka 599-8531, Japan.\\

\vspace{10mm}

$^*$ ishii@mtr.osakafu-u.ac.jp
\end{center}
\vspace{5mm}

\clearpage

\begin{figure}[t]
\begin{center}
\includegraphics[width=160mm]{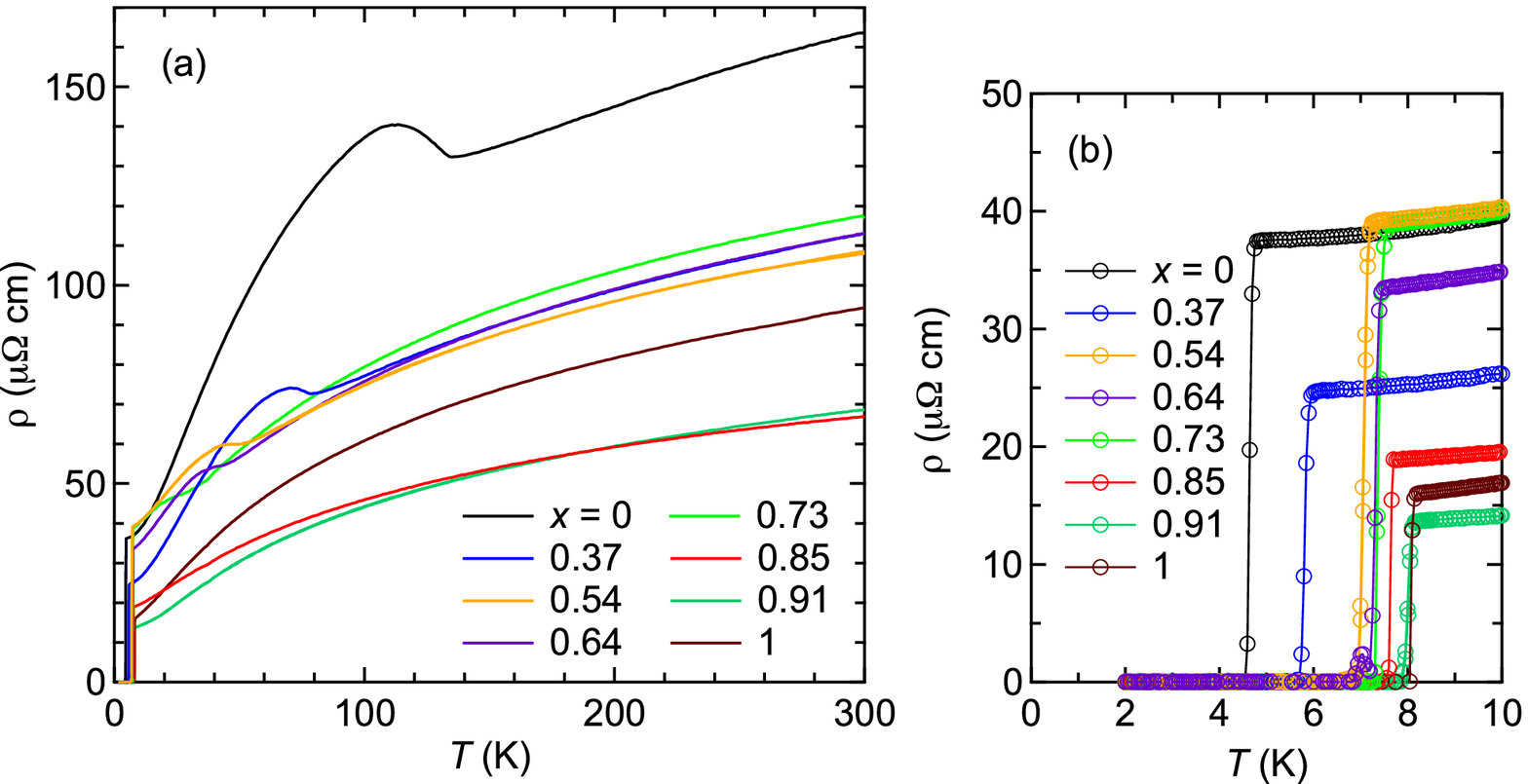}
\caption{\label{resistivity} 
(a) Electrical resistivity measured with alternating currents of 200--500 $\mu$A in a zero external magnetic field. (b) Enlarged view near the superconducting transition temperatures. 
$T_{\rm c}$ was defined as the temperature where zero resistivity was observed.
In the resistivity curves, the employed sweep rate was 0.1$^{\circ}$C/min near $T_{\rm c}$, where the heating process was completely in accord with the cooling process. 
Notably, reentrant behavior of superconductivity is observed in the $x=0.64$ sample. 
}
\end{center}
\end{figure}

\begin{table}[h]
\caption{
Parameters obtained by analyzing the electrical resistivity and specific heat data of (Sr$_{1-x}$Ca$_x$)$_3$Rh$_4$Sn$_{13}$. $T_{\rm c \; (\rho)}$ and $T_{\rm c \; (C_{\rm p})}$ are the superconducting transition temperatures obtained from electrical resistivity and specific heat measurements, respectively. $A$ and $\alpha$ are the temperature coefficient and the temperature exponent, respectively, in $\rho = \rho_0 + AT^{\alpha}$. Because the temperature exponents $\alpha$ for the samples of $x=$0--0.64 largely deviate from $\alpha=2$, the values of $A$ for these compositions are not listed.
}
\label{t1}
\begin{center}
\begin{tabular}{cccccccc}
\hline\hline
$x$ & $T_{\rm s}$ (K) & $T_{\rm c \; (\rho)}$ (K) & $T_{\rm c \; (C_{\rm p})}$ (K) & $\Delta / k_{\rm B}$ (K) & $2\Delta / k_{\rm B} T_{\rm c}$ & $A$ ($\mu \Omega$ cm K$^{-2}$) & $\alpha$ \\ \hline
0 & 135.1 & 4.5 & 4.4 & 7.88 & 3.61 & - & 3.74 \\
0.37 & 81.1 & 5.7 & 5.6 & 10.41 & 3.69 & - & 3.54 \\
0.54 & 51.1 & 6.9 & 6.3 & 12.84 & 4.06 & - & 2.98 \\ 
0.64 & 44.9 & 7.2 & 6.6 & 14.44 & 4.39 & - & 2.81 \\
0.73 & 29.1 & 7.3 & 6.9 & 16.27 & 4.71 & 1.25 $\times 10^{-2}$ & 2.28 \\
0.85 & - & 7.6 & 7.4 & 20.06 & 5.45 & 0.78 $\times 10^{-2}$ & 2.28 \\
0.91 & - & 7.8 & 7.5 & 20.68 & 5.53 & 1.08 $\times 10^{-2}$ & 2.09 \\ 
1 & - & 8.1 & 7.9 & 21.98 & 5.54 & 1.25 $\times 10^{-2}$ & 2.28 \\ \hline\hline
\end{tabular}
\end{center}
\end{table}

\clearpage

Specific heat measurements were performed for the samples of $x=0\sim1$ in zero and applied (6 T) external magnetic fields.
A superconducting jump is observed for each sample, and $T_{\rm c}$ is defined as the midpoint of each jump.

\begin{figure}[h]
\begin{center}
\includegraphics[width=160mm]{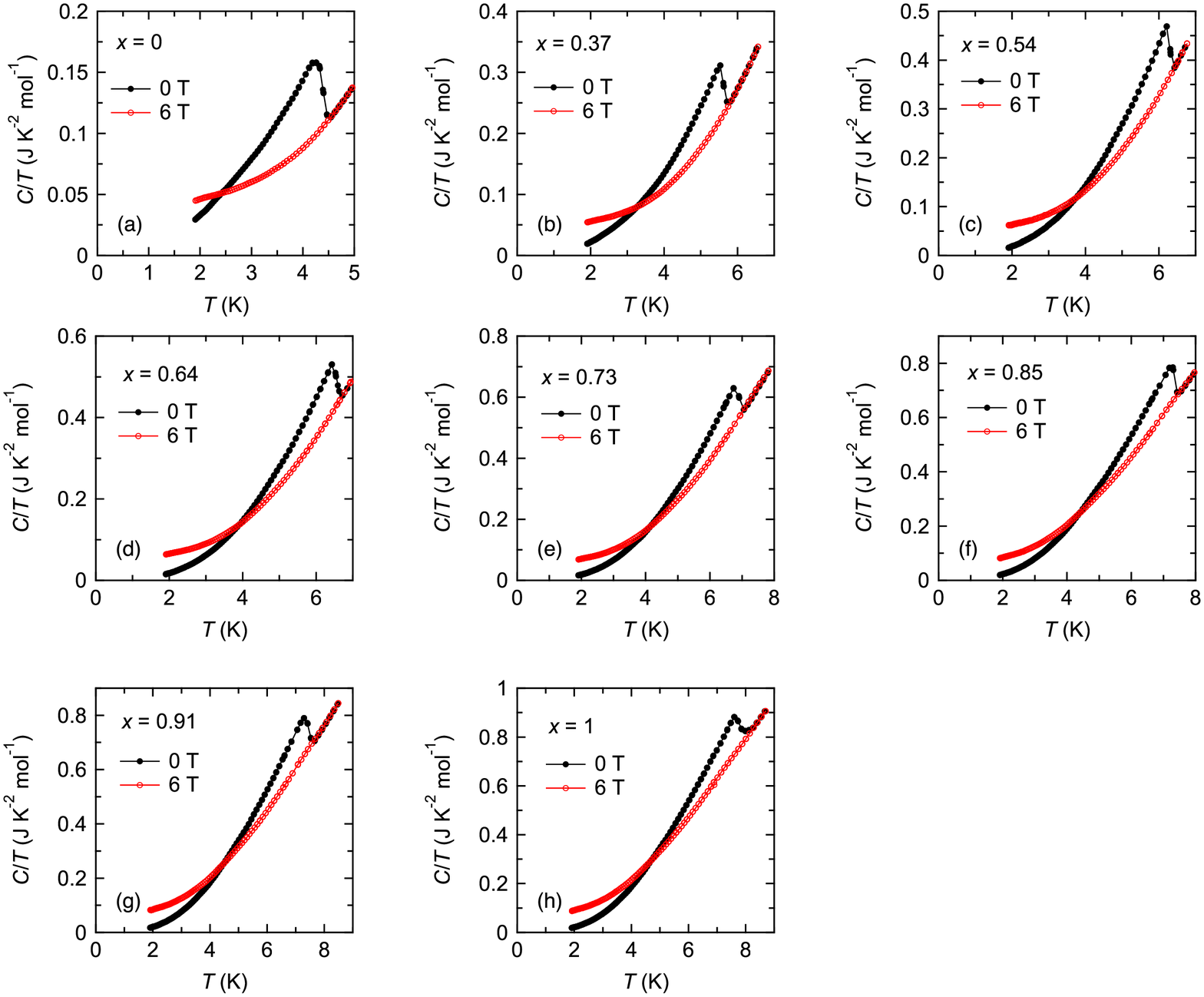}
\caption{\label{Cp_transition_each} 
Specific heat of $x=$0--1 samples measured in applied external fields of 0 T and 6 T. (a) $x = 0$, (b) 0.37, (c) 0.54, (d) 0.64, (e) 0.73, (f) 0.85, (g) 0.91, and (h) 1. 
}
\end{center}
\end{figure}

\clearpage

\begin{figure}[h]
\begin{center}
\includegraphics[width=160mm]{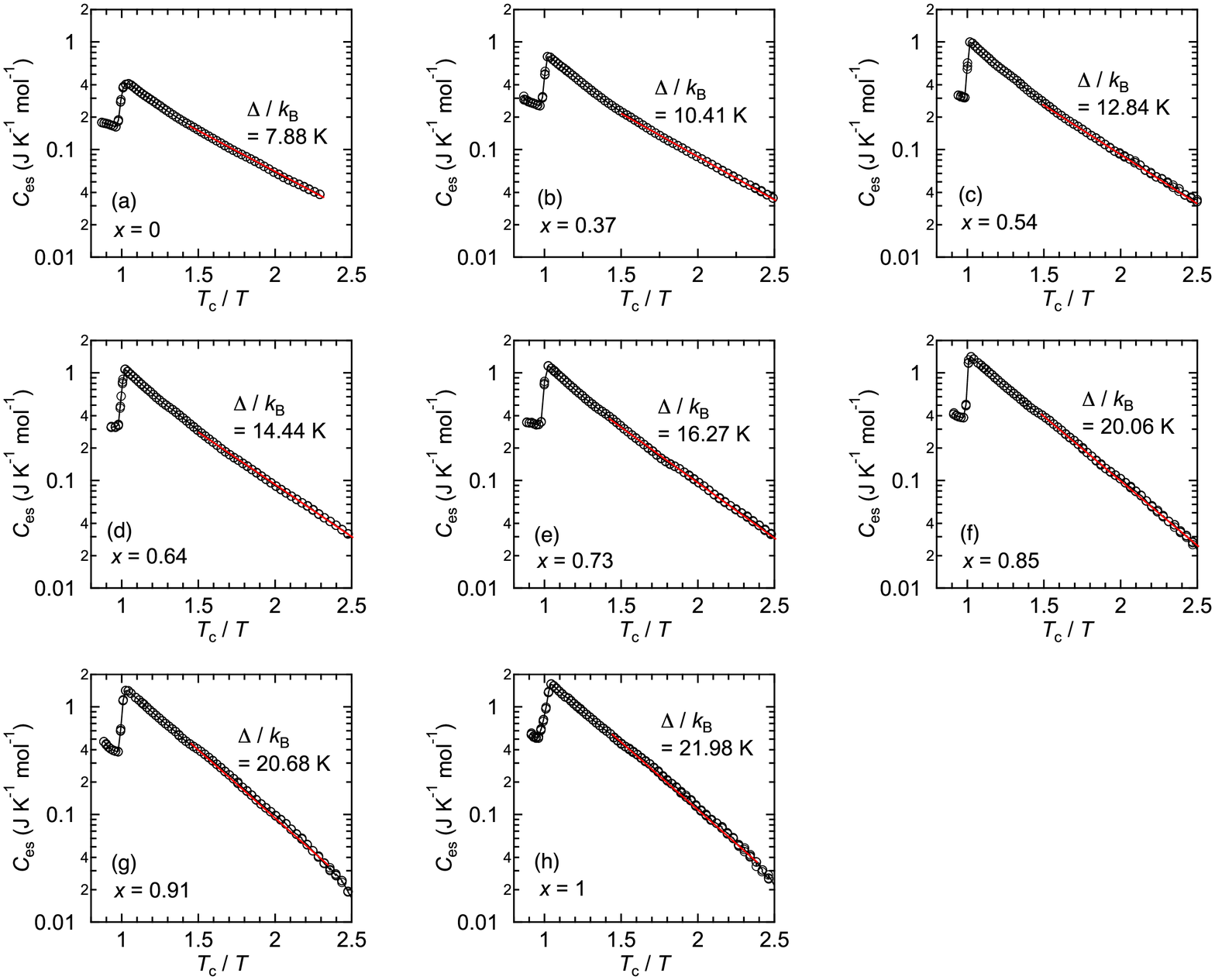}
\caption{\label{Ces_each} 
Electronic specific heat in the superconducting state, $C_{\rm es}$, for $x=$0--1 samples plotted as a function of $T_{\rm c} / T$. 
(a) $x = 0$, (b) 0.37, (c) 0.54, (d) 0.64, (e) 0.73, (f) 0.85, (g) 0.91, and (h) 1. 
}
\end{center}
\end{figure}

Electronic specific heat in the superconducting state, $C_{\rm es}$, was evaluated by subtracting the lattice contribution ($\beta T^3$) from the specific heat in the superconducting state. 
As shown in Fig. \ref{Ces_each}, $C_{\rm es}$ is well described as an exponential form of $B\exp(-\Delta / k_{\rm B} T)$ below $\sim$0.7 $T_{\rm c}$.
The superconducting gap, $\Delta$, was obtained as a slope of $\ln (C_{\rm es})$ vs. $T^{-1}$ and is summarized in Table S1 together with the reduced value $2\Delta / k_{\rm B} T_{\rm c}$.
The obtained values are in good agreement with the previous reports [S1, S2].

[S1] W. C. Yu, Y. W. Cheung, P.J. Saines, M. Imai, T. Matsumoto, C. Michioka, K. Yoshimura, and S. K. Goh, {\it Phys. Rev. Lett.} {\bf 115}, 207003 (2015).
[S2] H. Hayamizu, N. Kase, and J. Akimitsu, {\it J. Phys. Soc. Jpn.} {\bf 80}, SA114 (2011).